\documentstyle[prl,aps,preprint]{revtex}
\def\thru#1{\mathrel{\mathop{#1\!\!\!/}}}
\begin{document}
\tighten

\title{MATCHING PERTURBATIVE AND NON-PERTURBATIVE PHYSICS
WITH POWER ACCURACY IN HEAVY-QUARK EFFECTIVE THEORY\thanks
{This work was supported in part by funds provided by the U.S.
Department of Energy (DOE) under cooperative agreement No. DF-FC02-94ER40818.}}

\author{Xiangdong Ji}

\address{Center for Theoretical Physics \\
Laboratory for Nuclear Science \\
and Department of Physics \\
Massachusetts Institute of Technology \\
Cambridge, Massachusetts 02139 \\
{~}}

\date{MIT-CTP-2453 \hskip 1in  hep-ph/9507322 \hskip 1in July 1995}

\maketitle

\begin{abstract}
I discuss a scheme to match perturbative and non-perturbative
physics with power accuracy in the heavy-quark effective
theory. I elaborate on two important aspects of the scheme: 1)
a multi-loop subtraction of soft contributions from
the perturbation series in the pole mass, 2)
a perturbative regularization of
the linearly-divergent heavy-quark self-energy in
the lattice formulation of the heavy-quark effective
theory.
\end{abstract}

\pacs{xxxxxx}

\narrowtext
In the last few years, the heavy-quark effective theory, HQET,
has been widely used to study hadron physics involving
heavy flavors. The HQET is essentially
based on the concept of
scale separation---the large-momentum-scale physics associated with
the heavy-quark mass can be factorized and calculated in
perturbation theory, and the remaining low-momentum-scale,
non-perturbative physics can be summarized in the heavy-quark
effective lagrangian. Since the number of papers on the subject
is enormous, I refer the reader to
review articles for relevant references \cite{neubert}.

It was generally accepted that the heavy-quark mass
used in the heavy-quark expansion should be
the pole mass, defined according to the single-particle
pole in the perturbative heavy-quark propagator\cite{novikov1}.
Recently, however, Bigi, Shifman, Uraltsev, and Vainshtein (BSUV)
\cite{bigi} and Beneke and Brown\cite{beneke}
pointed out that the pole mass
is intrinsically ambiguous at the order of
$\Lambda_{\rm QCD}$ and thus, to this degree,
the heavy-quark expansion is not unique. The discovery
is consistent with an earlier observation by Falk, Neubert
and Luke \cite{falk} that a residual mass term  $-\delta m_0 \bar h_v h_v$
is generally present in the heavy-quark effective lagrangian.
It is also consistent with the finding by
Maiani, Martinelli and Sachrajda \cite{maiani} that the lattice
formulation of the HQET at order $1/m_Q$
is plagued by the divergences in inverse powers of the
lattice spacing $a$.

Given the freedom of constructing the HQET, it is important
to find a scheme to match perturbative
and non-perturbative calculations with power accuracy. Recently,
two approaches have appeared in the literature to tackle
this issue. After arguing that a perturbative
subtraction of power divergences is not viable,
Martinelli and Sachrajda \cite{martinelli1}
proposed to use non-perturbative renormalization
conditions to define power-divergent operators
\cite{martinelli2}.
They provided examples to show how it works in the lattice
simulations of the HQET. In their approach,
the cancellation of the infrared renormalons in perturbation
series involves non-perturbative physics. On the other hand,
BSUV \cite{bigi} took Novikov et al.'s approach to Wilson's
operator-product expansion (OPE)\cite{novikov}.
According to that, the OPE provides
a separation of physics above and below
a certain momentum scale $\mu$. The low-momentum
physics is taken into account by the non-perturbative
matrix elements and the high-momentum physics is included
in the coefficient functions. In their paper, BSUV showed
explicitly how this can be done at
one-loop level.

In this paper I follow the proposal by
BSUV and aim to complete the scheme in two
important aspects:
1). Outline a procedure for the multi-loop subtraction of the
soft contributions present in the perturbation
series in the pole mass. The subtraction can be
calculated perturbatively using the heavy-quark effective
lagrangian in the dimensional regularization and minimal
subtraction scheme ($\overline{\rm MS}$). The approach
works also for other perturbation series in the HQET.
 2). Show how the non-perturbative corrections
may be computed, consistently with the above
perturbative calculation, in the lattice formulation of
the HQET. This involves matching the
non-perturbative quantities in the
$\overline{\rm MS}$ and lattice schemes and
formulating the renormalization conditions
for power-divergence quantities in perturbative
theory. The discussion follows closely a recent paper
on Wilson's expansion with power accuracy
by the present author\cite{ji}.

I start with the heavy-quark expansion for the
projected inverse heavy-quark propagator in the full theory, as
studied by Beneke and Braun \cite{beneke},
\begin{equation}
    S^{-1}_P(k, m_Q) =
      m_Q - m_{\rm pole}(m(\mu), \alpha_s(\mu))
     + C({m_Q\over \mu}, \alpha_s(\mu))
      S_{\rm eff}^{-1}(v\cdot k, \mu)
     +{\cal O}({(v\cdot k)^2\over m_Q^2}, {k_\perp^2\over m_Q^2}) \ ,
\end{equation}
where $m_Q$, to be specified shortly,
is the heavy-quark mass defining the expansion,
$v$ is the velocity of the
heavy quark, and $k$ is the residual momentum.
Their relation with the total momentum of the heavy quark, $P$, is
\begin{equation}
      P = m_Q v + k \ .
\end{equation}
The pole mass $m_{\rm pole}$ is a function of
the renormalized $\overline{\rm MS}$ mass $m(\mu)$.
It is also a perturbation series in $\alpha_s$,
plagued by the infrared renormalon at $b=2\pi/\beta_0$ in
the Borel plane\cite{bigi,beneke}, where $\beta_0
=11-2N_f/3$. $S^{-1}_{\rm eff}$ is a propagator
defined in the effective theory with the effective lagrangian
\begin{equation}
   { \cal L}_{\rm eff} = \bar h_v iv\cdot D h_v + {
    \cal L}_{\rm light} \ ,
\label{lag}
\end{equation}
in the $\overline{\rm MS}$ scheme, where $h_v$ is the
heavy-quark effective field and ${\cal L}_{\rm light}$
is the lagrangian for light quarks and gluons. $S^{-1}_{\rm eff}$
is a non-perturbative quantity with the ultraviolet renormalon at $
b=2\pi/\beta_0$ in the Borel plane\cite{beneke}.

It was suggested in Refs. \cite{bigi,beneke} that to
cancel the infrared renormalon, one
may subtract a residual mass $\delta m(\Lambda)$
of order $\Lambda_{\rm QCD}$ from
the pole mass $m_{\rm pole}$ and choose,
\begin{equation}
     m_Q(\Lambda) = m_{\rm pole} - \delta m(\Lambda) +... \ ,
\label{mq}
\end{equation}
as the expansion parameter,
where the ellipsis represents higher-order
terms in the heavy-quark mass. Apart from the role of
cancelling the renormalon,
$\delta m(\Lambda)$ is otherwise arbitrary. In Ref. \cite{bigi},
$\delta m(\Lambda)$ is identified as the Coulomb
energy of the heavy quark below the momentum
scale $\Lambda$. If so, the expansion parameter
$m_Q(\Lambda)$ contains only the physics above the scale
$\Lambda$ and is entirely perturbative if $\Lambda\gg \Lambda_{\rm QCD}$.
The $\Lambda$ dependence in $m_Q(\Lambda)$ shows explicitly
 that the
construction of the heavy-quark expansion is not unique.
The scale $\Lambda$ needs not to be the same as the renormalization scale
$\mu$ introduced in the full theory.

To determine the subtraction to all orders, I consider
the heavy-quark propagator
$S^{-1}_{\rm eff}(v\cdot k)$ in perturbation theory.
[I will neglect the generic renormalization scale $\mu$ below.]
In dimensional regularization, the propagator satisfies
the renormalization condition,
\begin{equation}
S^{-1}_{\rm eff}(v\cdot k=0, ~\overline{\rm MS})|_{\rm pert} = 0 ,
\label{ren}
\end{equation}
i.e., the mass shift vanishes.
The argument for this is simple:
The effective lagrangian ${\cal L}_{\rm eff}$
contains no mass scale in the chiral limit. Because
$S^{-1}_{\rm eff}(v\cdot k=0)|_{\rm pert}$
has the dimension of a mass, it must diverge
linearly. All power
divergent integrals are taken to be zero
in dimensional regularization.

However, the conclusion is deceptive.
The self-energy of the heavy quark
contains both soft and hard contributions, and they
cannot cancel each other on the physical ground. For definiteness,
let us define the hard and soft contributions as follows.
Denote a generic light-quark or gluon propagator
by $D(k)$. Split the propagator into two parts according
to virtuality $k^2$ of the particle,
\begin{equation}
      D(k) = D(k)\theta(k^2>\Lambda^2) + D(k)\theta(k^2<\Lambda^2)\ .
\end{equation}
Then the contribution to the heavy-quark self-energy from
a Feynman diagram with all propagators replaced by
$D(k)\theta(k^2>\Lambda^2)$  is defined as hard.
The complementary part, denoted as
$\Delta \Sigma$, is soft,
\begin{equation}
    \Delta \Sigma(\Lambda)
     = S^{-1}_{\rm eff}(v\cdot k=0)|_{\rm pert}
      - S^{-1}_{\rm eff}(v\cdot k=0)|_{\rm pert}(
    {\rm all}~ k^2>\Lambda^2)\ .
\label{sub1}
\end{equation}
It is easy to see that $\Delta \Sigma$ does not have
any linear divergences. All
logarithmic divergences in it can be regulated dimensionally and
subtracted minimally. Therefore we define the residual
mass,
\begin{equation}
    \delta m(\Lambda) = C(
     \alpha_s) \Delta \Sigma(\Lambda)\ ,
\label{sub2}
\end{equation}
which is a perturbation series in $\alpha_s$,
$\Lambda\sum_{n=1}b_n \alpha_s^n$.
The one-loop calculation in Ref. \cite{bigi}
gives $c_1 = 2/3$. Multi-loop subtraction
can now be routinely calculated according to
Eqs. (\ref{sub1}) and (\ref{sub2}). Since $\delta m(\Lambda)$ contains
all the soft contribution, it has the infrared renormalon
at $b=2\pi/\beta_0$.

Following Refs. \cite{bigi,beneke}, I reorganize Eq. (1) as follows,
\begin{equation}
    S^{-1}_P(k, m_Q) =
      m_Q - [m_{\rm pole}-\delta m(\Lambda)]
     + C( \alpha_s) \left[
      S_{\rm eff}^{-1}(v\cdot k,~ \overline{\rm MS}) - \Delta \Sigma(\Lambda)
     \right]
     + ... \ .
\label{exp1}
\end{equation}
The infrared renormalon in $\delta m(\Lambda)$ cancels
both the infrared renormalon in $ m_{\rm pole}$
and the ultraviolet renormalon in $S_{\rm eff}$,
rendering $m_{\rm pole} - \delta m(\Lambda)$
and $S_{\rm eff}^{-1} - \Delta \Sigma(\Lambda)$ well-defined.
However, both quantities now depend on the separation
scale $\Lambda$.

I now turn to the non-perturbative part of the expansion
in the lattice formulation of QCD, because at present the lattice
provides the only formalism to calculate non-perturbative
physics from the fundamental lagrangian. In a
lattice, the discretized version of ${\cal L}_{\rm eff}$ produces
a linearly-divergent quark self-energy. To calculate
$S_{\rm eff}^{-1}$ in the $\overline{\rm MS}$-scheme,
we first match it with
the lattice propagator,
\begin{equation}
   S_{\rm eff}^{-1}(v\cdot k, \overline{\rm MS})
    = Z(\mu a, \alpha_s) \left[S_{\rm eff}^{-1}(v\cdot k,~ {\rm latt})
          - S_{\rm eff}^{-1}(0, ~{\rm latt})_{\rm pert}\right]\ ,
\label{newexp}
\end{equation}
where the second term, calculated in the lattice
perturbation theory, diverges like $1/a$
and serves to cancel the linear divergence in
the first term. $Z$ is a renormalization constant,
calculable in perturbation theory\cite{ji1}.
Alternatively, one can add a residual
mass term $-\delta m_0\bar h_vh_v$ to the original lagrangian
in Eq. (\ref{lag}), where $\delta m_0$ is fixed by
the perturbative renormalization condition,
\begin{equation}
 S_{\rm eff}^{-1}(0,  ~{\rm latt})|_{\rm pert. } = 0 \ .
\end{equation}
The propagator calculated in the new effective
theory has no linear divergences. The new propagator
replaces the two terms in the right-hand side of
Eq. (\ref{newexp}).
However, the residual mass thus determined,
$\delta m_0 = 1/a\sum_n c_n \alpha_s^n$,
has the infrared renormalon at $b=2\pi/\beta_0$.

There is yet another way to define the
effective theory on the lattice, which is free
of both the linear divergence and the infrared
renormalon. Define a new $\delta m_0$ with a subtraction
of another series,
\begin{equation}
    \delta m_0 = 1/a\sum_{n=1}^\infty c_n \alpha_s^n
    - \Lambda\sum_{n=1}^\infty d_n \alpha_s^n \ ,
\end{equation}
where $d_n$ is adjusted so that
the following renormalization condition is satisfied,
\begin{equation}
 S_{\rm eff}^{-1}(0, ~{\rm latt})|_{\rm pert } = -Z^{-1}
 \Delta \Sigma(\Lambda) \ .
\label{rc}
\end{equation}
Then $\delta m_0$
is free of the infrared renormalon, although it now depends on
the separation scale $\Lambda$. The modified effective lagrangian,
\begin{equation}
    {\cal L'}_{\rm eff}(\Lambda) =
    {\cal L}_{\rm eff} - \delta m_0(a, \Lambda)\bar h_v h_v\ ,
\end{equation}
is also $\Lambda$ dependent, as discussed in Refs. \cite{beneke,falk}.
The non-perturbative
heavy-quark propagators $S_{\rm eff}^{-1}(v\cdot k, \Lambda)
$ in this theory, apart from the renormalization factor $Z$,
gives directly $S_{\rm eff}^{-1}(v\cdot k, \overline{\rm MS}) -
\Delta \Sigma(\Lambda)$
in Eq. (\ref{exp1}).
Let me emphasize
again that the residual $\delta m_0$ is an entirely
perturbative quantity.

Using the expansion parameter defined in Eq. (\ref{mq}),
I have,
\begin{equation}
    S^{-1}_P(k, m_Q) = C(\alpha_s)
     Z S_{\rm eff}^{-1}(v\cdot k, \Lambda)
+ ... \ .
\label{exp2}
\end{equation}
It is possible to find a special value of $\Lambda=\Lambda_0$ such that,
\begin{equation}
     S_{\rm eff}^{-1}(0, \Lambda_0) = 0\ .
\end{equation}
The expansion parameter $m_Q^S$ define at this $\Lambda_0$
coincides with the subtracted pole mass
introduced by Martinelli and Sachrajda.
In some sense, this special mass is the physical
pole mass because the propagator in the full theory
has a pole at $\thru p = m_Q^S$, as is clear
from Eq. (\ref{exp2}). $m_Q^S$ is
independent of the details of the effective
theory. It would be nice to check that the two formulations of
the lattice renormalization conditions give the same
$m_Q^S$.

Both the subtracted pole mass and the physical mass
of a pseudo-scalar meson with the heavy quark
can be expanded in the $m_Q(\Lambda)$,
\begin{eqnarray}
    m_Q^S && = m_Q(\Lambda) + C(\alpha_s)S^{-1}_{\rm eff}(0,\Lambda) + ... \ ,
\nonumber \\
    m_H &&= m_Q(\Lambda) + C'(\alpha_s)\bar \Lambda(\Lambda) + ... \ .
\end{eqnarray}
$\bar \Lambda$ is the mass of the pseudo-scalar
meson calculated in the
lattice effective theory with the renormalization condition
in Eq. (\ref{rc}).
$C$ and $C'$ are the coefficient functions
calculable in perturbation theory.
$\Lambda$ dependence is explicit in different orders of
$m_Q$ in the above equations, however, it cancels in
the physical quantities.

In conclusion, I have discussed in this paper
the heavy-quark effective
theory in the approach of BSUV. I gave a multi-loop subtraction
formula for calculating the scale-dependent heavy-quark mass
$m_Q(\Lambda)$. I outlined a perturbative regularization
of the linear divergences in the heavy-quark self-energy
in the lattice QCD. With these crucial ingredients,
power corrections to the results of the leading-order
HQET can be practically calculated.

\acknowledgments
I thank M. Beneke for discussions about
renormalons in the HQET.

\end{document}